\documentclass{iaus}
\usepackage{graphicx}

\title[IAUS 262.~~A large sample of LSBGs from SDSS] %% give here short title %%
{The properties of a large sample of low surface brightness galaxies from SDSS}

\author[Liang et al.]   %% give here short author list %%
{Y. C. Liang$^1$, G. H. Zhong$^{1,2}$, X. Y. Chen$^{1,2}$, D. Gao$^{1,2}$, 
F. Hammer$^3$, F. S. Liu$^{4,1}$,
J. Y. Hu$^1$, L. C. Deng$^1$, B. Zhang$^{5,1}$
}

\affiliation{$^1$National Astronomical Observatories, Chinese Academy of
Sciences, A 20 Datun Road Chaoyang District, Beijing 100012, China;
 email: {\tt ycliang@nao.cas.cn}
\\[\affilskip]$^2$Graduate School of the Chinese Academy of Sciences, Beijing
100049, China
\\[\affilskip]$^3$GEPI, Observatoire de Paris-Meudon, 92195 Meudon, France
\\[\affilskip]$^4$College of Physics Science and Technology, Shenyang Normal Univ., Shenyang
110034, China
\\[\affilskip]$^5$Department of Physicals, Hebei Normal University,
Shijiazhuang 050016, China  
}

\pubyear{2009}
\volume{262}  %% insert here IAU Symposium No.
\pagerange{1--4}
\jname{Stellar Populations: Planning for the Next Decade}
\editors{G. Bruzual \& S. Charlot, eds.}
\begin{document}

\maketitle

\begin{abstract}
A large sample of low surface brightness (LSB) disk galaxies is selected 
from SDSS with 
B-band central surface brightness $\mu_0$(B) from 22 to 24.5 mag arcsec$^{-2}$.
Some of their properties are studied, such as 
magnitudes, surface brightness, scalelengths, colors, 
metallicities, stellar populations, stellar masses
and multiwavelength SEDs from UV to IR etc. 
These properties of LSB galaxies have been compared with those of the 
galaxies with higher surface brightnesses. Then we check the variations
of these properties following surface brightness.   
\keywords{galaxies: abundances, galaxies: evolution, 
 galaxies: spiral, galaxies: starburst}
\end{abstract}

\firstsection % if your document starts with a section,
\section{Introduction}

Low Surface Brightness Galaxies (LSBGs) are important populations
in galaxy field. However, their  contributions to galaxy
population have been underestimated for a long time since they
are hard to find owing to their faintness compared  with the
night sky.  An initial quantitative study was made by Freeman
(1970), who noticed that the central surface brightness of their
28 out of 36 disc  galaxies fell within a rather  narrow range,
$\mu_0$(B)=21.65$\pm$0.3 mag arcsec$^{-2}$. This could be caused
by selection effects (Disney 1976).

Since then, many efforts have been made to search for large
number of LSBGs  from surveys (Bothun \& Impey 1997; Impey \&
Bothun 1997;  Zhong et al. 2008 and references therein).  One of
the important ones is the APM survey, i.e., Impey et al. adopted
the Automated Plate Measuring (APM) mechanism to scan UK Schmidt
plates and  discovered 693 LSBGs  which forms the most extensive
catalog of LSBGs to that date (Impey et al. 1996). O'Neil et al.
(1997, in ``Texas survey")  firstly found the red LSBG
populations. Monnier Ragaigne et al. (2003) selected a sample of
about 3800 LSBGs from the all-sky near-infrared 2MASS survey, and
then made HI observations for a sub-sample. 

The modern digital sky survey, such as the Sloan Digital Sky
Survey (SDSS),  certainly provide a wonderful chance for us to find
much more LSBGs.  Kniazev et al. (2004) developed a
method to search for LSBGs from SDSS images and used the APM
sample to test their method. They recover 87 same objects and 42
new LSBGs. Recently, our group (Zhong et al. 2008) successfully
select a large sample of LSBGs from SDSS-DR4 main galaxy sample,
which includes 12,282 nearly face-on disk galaxies with low
surface brightness $\mu_0$(B)$\geq$ 22 mag arcsec$^{-2}$, and
another 18,051 high surface brightness galaxies (HSBGs) with 
$\mu_0$(B)$<$22 mag arcsec$^{-2}$ are also selected for
comparisons. This will be a very efficient sample  to study the
properties of LSBGs, and to check the variation of these
properties with surface brightnesses if there exists.  We have
studied some properties of this large sample. Zhong et al. (2008)
presented the sample selection criteria, and their magnitudes,
surface brightness, scalelengths and colors etc.
The 
metallicities (Liang et al. 2009, in preparation) and  stellar
populations (Chen et al. 2009, in preparation) are also studied
from optical spectra, i.e. the emission lines, continua and
absorption lines. In addition, the modern digital sky surveys
have been made in  wide wavelength ranges, such as GALEX-UV,
SDSS-optical, 2MASS-NIR, IRAS-IR etc. These allow us to construct
the multi-wavelength SEDs for our sample galaxies  (Gao et al.
2009, in preparation), which are very useful and efficient to
study their star formation history.
In this paper, we summarize the interesting results we 
obtained for this large sample of  galaxies, which will be
presented in detail in the  series work mentioned above.
Especially, we try to  check the varying trends of their
properties following surface brightness. 
A cosmological model with $H_0$ = 70 km s$^{-1}$ Mpc$^{-1}$,
$\Omega_M$ = 0.3  and $\Omega_\lambda$ = 0.7 is adopted in this
paper.
\section{The sample, the correlations between $M_B$ and log$h$, log$D$ and log$h$}
 
We select 30,333 nearly face-on disk galaxies from 
the SDSS-DR4 main galaxy sample. They have been
estimated the reliable surface brightness values 
and minimize the effect of dust extinction inside the galaxies.
The selection criteria are given below.
 
 \begin{enumerate}

\item  $fracDev_r$ $<$ 0.25 ($fracDev_r$ indicates the fraction of luminosity
contributed by the de Vaucouleurs profile relative to exponential
profile in the $r$-band, and this low value for limit can also minimize the effect
of bulge light on the disk galaxies); 
\item $b/a$ $>$ 0.75
(for nearly face-on galaxies, where $a$ and $b$ are the semi-major and
semi-minor axes of the fitted exponential disk respectively); 
\item $M_B$ $<$ -18 (this 
excludes the few (6\%) dwarf galaxies contained in the sample). 

 \end{enumerate}

Taking $\mu_0$(B)=22 mag arcsec$^{-2}$ as criterion,  12,282
objects are confirmed as LSBGs and other 18,051 objects are
confirmed as HSBGs.  The properties of this large sample of LSBGs
have been studied carefully by Zhong et al. (2008). Specially, 
these LSBGs show clear correlations between $M_B$ and log$h$, and
between log$D$ and log$h$, which mean that the brighter galaxies tend
to have larger scalelengths and the galaxies with larger
scalelengths have the benefit  to be detected and observed in the
more distant universe.
 
 Furthermore, both of the LSBGs and HSBGs can be  divided into
two sub-groups following $\mu_0$(B), which will allow us to study
their properties in a sequence of surface brightness.  The
criteria for $\mu_0(B)$ (in mag arcsec$^{-2}$) are following
McGaugh (1996): 
sLSBGs (2,815) with 22.75-24.5,
ISBGs (9,467) with 22.0-22.75,
sHSBGs (10,989) with 21.25-22.0,
VHSBGs (7,062) with $<$21.25.
Fig.1a shows the relations of $M_B$ vs. log$h$  for the samples
in these four sub-groups.  Fig.1b shows that for log$D$ vs.
log$h$. All the four sub-groups show clear correlations. However,
the comparisons among these four show that, at a given disk
scalelength, the galaxies having higher $\mu_0$(B) are  brighter,
and could be observed at a farther distance.  There are 
gradually varying trends following surface brightnesses of the
galaxies.
\section{The metallicities}
 
Metallicity is an important parameter of galaxy and a good tracer
for their evolution history. The metallicities of this large
sample of nearly face-on disk galaxies  can be estimated from
optical emission lines. To select the target galaxies with good
quality spectroscopic observations, we adopt the criteria below
(following Liang et al. 2006 and Tremonti et al. 2004). 

\begin{enumerate}

\item Cross correlating 
between our sample and the SDSS-DR4 spectroscopic database.
 
 \item  The fluxes of the emission lines of the galaxies have been measured for 
 [O~{\sc ii}], H$\beta$, [O~{\sc iii}], H$\alpha$, [N~{\sc ii}],
 and the line H$\beta$, H$\alpha$, and [N~{\sc ii}]6584 are
 detected with S/N ratios greater than 5$\sigma$. Then the selected sample in
 the four sub-groups are 1,364 in sLSBGs, 6,055 in ISBGs, 9,107 in
 sHSBGs are 6,231 in VHSBGs.
 
\item Selecting the star-forming galaxies 
by using the diagnostic diagram of [N~{\sc ii}]/H$\alpha$ vs.
 [O~{\sc iii}]/H$\beta$ following Kauffmann et al. (2003a). Now we
 obtain
 1,299 of sLSBGs,
 5,551 of ISBGs, 
 8,310 of sHSBGs, and 
 5,872 of VHSBGs.
 This also means that the AGN fractions of these sub-groups are
quite small.
 
 \end{enumerate}

 The  12+log(O/H) abundances of these star-forming galaxies are
taken from  the MPA/ JHU database (Tremonti et al. 2004). The
median values in the four sub-groups are 8.77 for sLSBGs,  8.94
for ISBGs, 9.03 for sHSBGs, and 9.06 for VHSBGs. This means that
the galaxies  have lower surface brightness generally have lower
metallicity. 

 Fig.2a shows the relations of 12+log(O/H) vs. $\mu_0(B)$ of the
 star-forming galaxies in the four sub-groups. The median values
 in each of the bins of 0.2 in $\mu_0(B)$ are also given (the big
 squares), as well as the three vertical long-dashed lines at
 $\mu_0(B)$=21.25, 22.0 and 22.75 to mark the ranges of $\mu_0(B)$
 for the four sub-groups. A general varying trend shows that,
 for the sLSBGs, ISBGs and sHSBGs, the galaxies with lower surface
 brightness have lower metallicity. We further obtain the
 stellar masses of these sample galaxies  from the MPA/JHU
 database (Kauffmann et al. 2003b), and present their relations
 with  $\mu_0(B)$ in Fig.2b. It shows that, for the sLSBGs, ISBGs
 and sHSBGs,  the galaxies with lower surface brightness have
 smaller stellar mass generally. These are in agreement with the
 stellar mass-metallicity relations  of the galaxies.
\begin{figure}
\centering
\input epsf
\includegraphics [width=5.5cm, height=4.5cm]{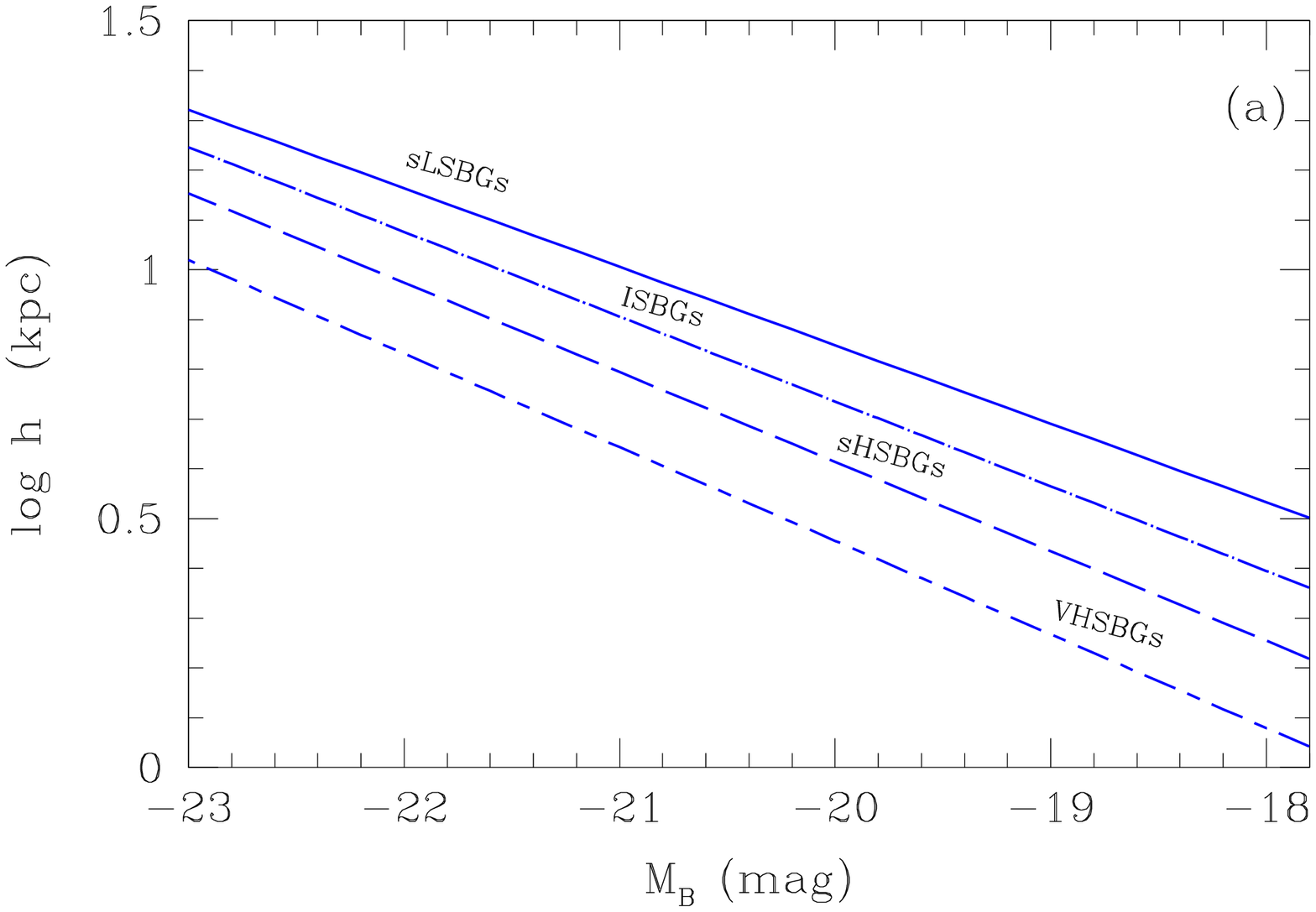} 
\includegraphics [width=5.5cm, height=4.5cm]{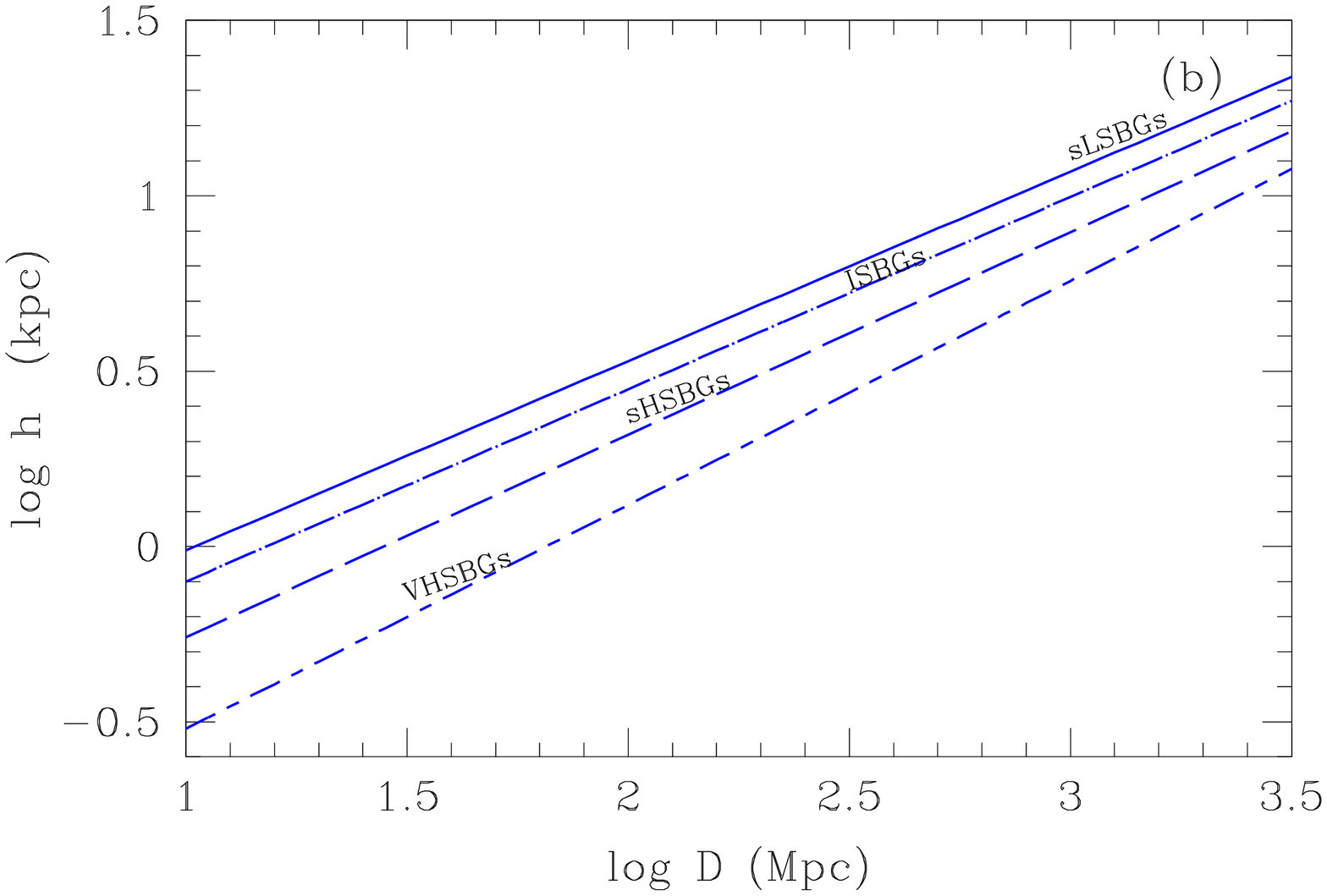} 
\vspace{-1.6cm}
\caption{(a) Correlation of disk scale length and luminosity for 
the whole sample galaxies with $\mu_0(B)$ from 18.0-24.5 mag arcsec$^{-2}$,
(b) correlation of disk scale length and distances for them: 
sLSBGs (24.5-22.75, the solid line); ISBGs (22.75-22.0, the dot-long dash line);
sHSBGs (22.0-21.25, the long-dash line); 
VHSBGs ($<$21.25, the short-long dash line).
      }
\label{fig.1}
\end{figure}

\begin{figure} 
\centering
\includegraphics [width=5.5cm, height=4.5cm]{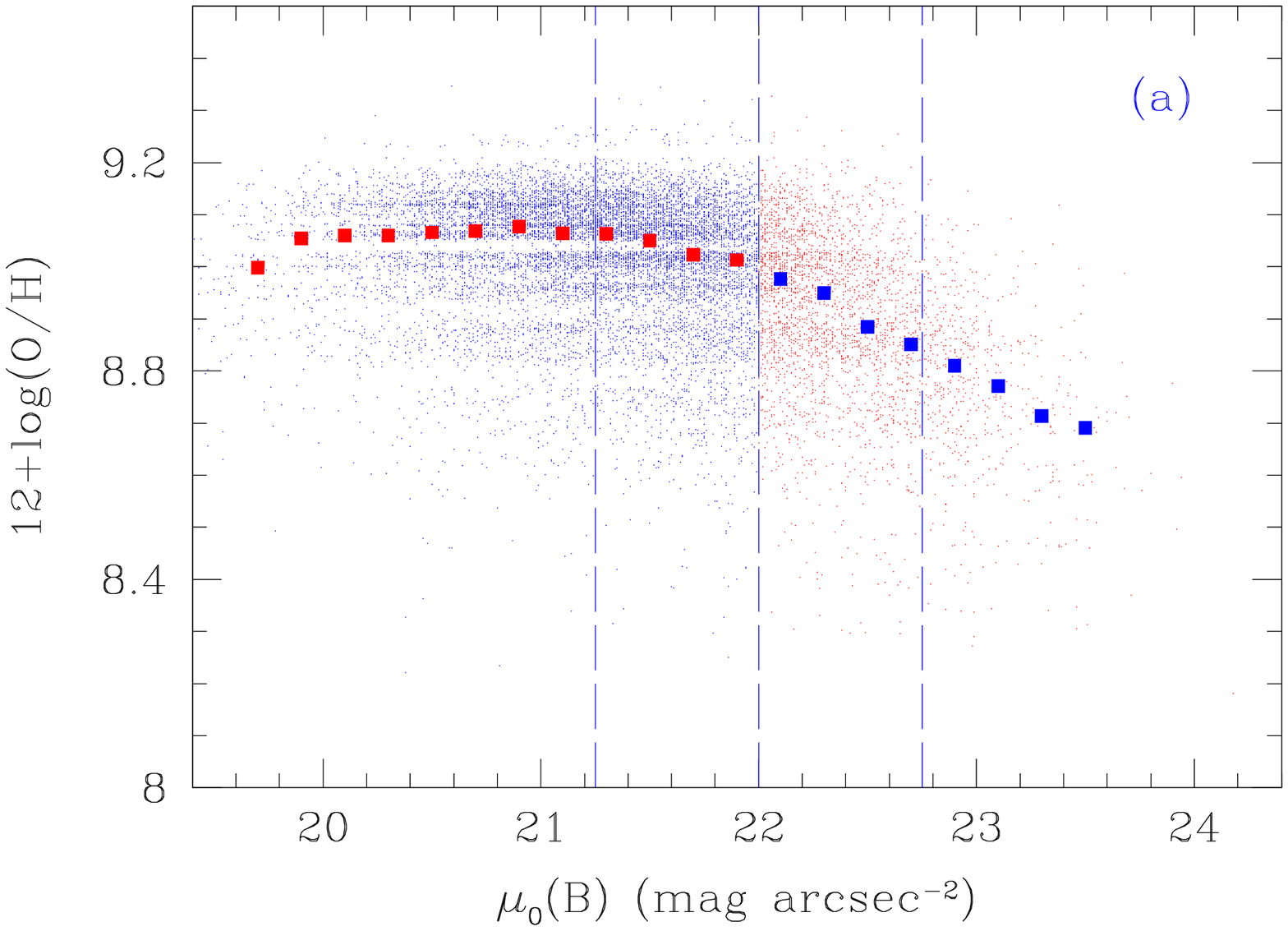} 
\includegraphics [width=5.5cm, height=4.5cm]{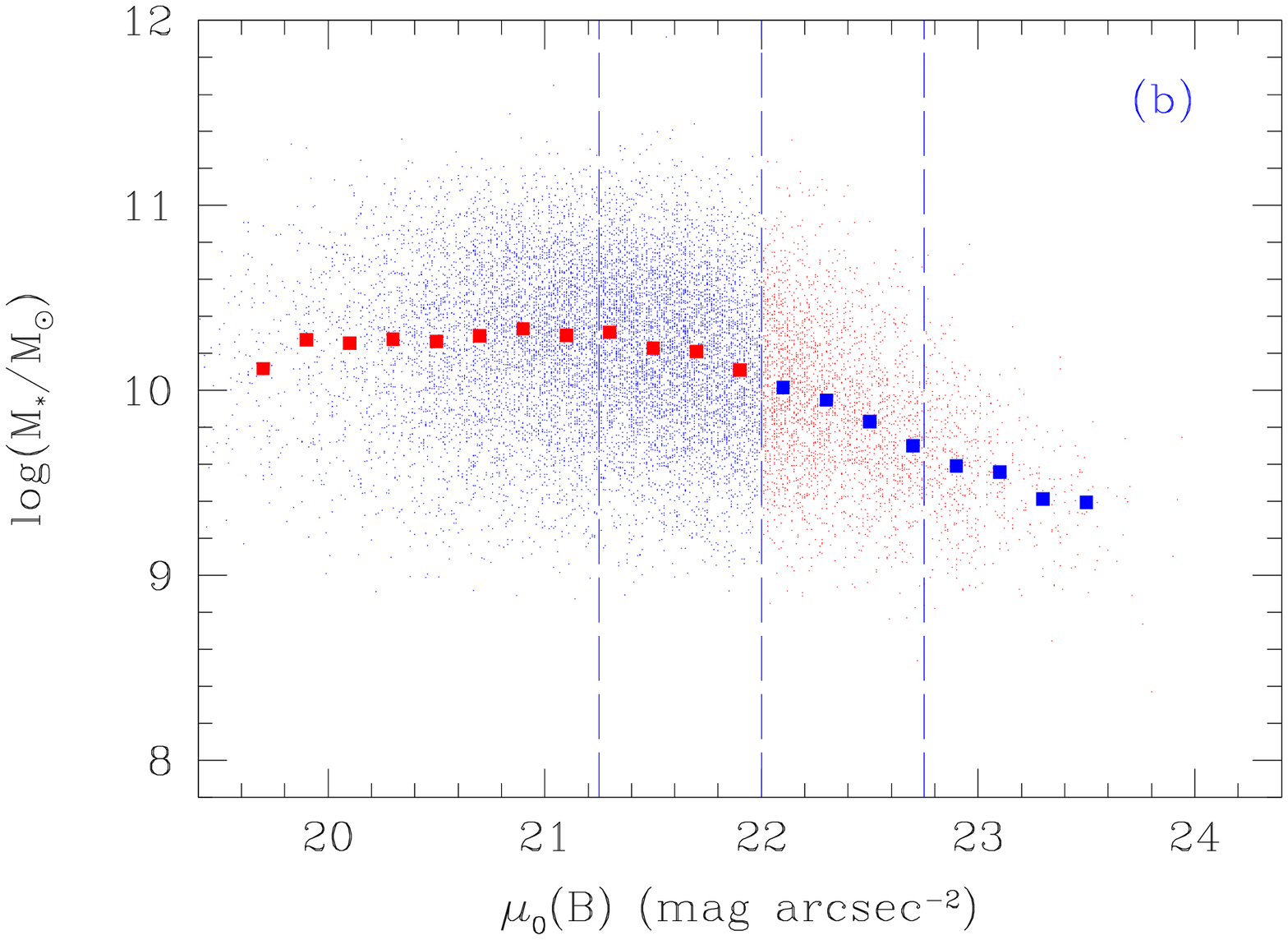} 
\vspace{-1.4cm}
\caption {(a) The relations of 12+log(O/H) vs. $\mu_0(B)$    for
the star-forming galaxies, (b) the relations of stellar mass vs.
$\mu_0(B)$  of them. The median values in each of the bins of 0.2
in $\mu_0(B)$ are also given as the big squares. The three
vertical long-dashed lines at 22.75, 22.0 and 21.25 mark the
ranges of $\mu_0(B)$ for the four sub-groups of sLSBGs, ISBGs,
sHSBGs, VHSBGs.}
\label{fig.oh.mb.mu}
\end{figure}
\section{The stellar populations and multiwavelength SEDs}

We study the stellar populations of this sample of star-forming
galaxies by using  the software STARLIGHT (Cid Fernandes et al.
2005; Chen et al. 2009) to fit the optical spectral absorptions
and continua.   The template for spectral synthesis analysis are
the simple stellar  populations from Bruzual \& Charlot (2003).   We
combined the spectra of all the objects to be one  spectrum in
each of the sub-samples, then we synthesize these combined
spectra.  The results show that the importance of young
populations increases from  sLSBGs to ISBGs, sHSBGs, and to VHSBGs
(more details in Chen et al. 2009, in preparation). 

Since the multiwavelength modern digital sky surveys have observed
and released a very large sample of objects to public,  such as
GALEX-UV, SDSS-optical, 2MASS-NIR and IRAS-IR (the Spitzer/SWIRE
survey also release their IR data, but in a much smaller sky
coverage).  Therefore, we cross correlate our such large sample of
nearly face-on disk galaxies selected from optical with the
observations at UV, NIR and IR, and then construct their
multiwavelength SEDs. We extend the SDSS-DR4 to the DR7 main
galaxy sample here, which   expands to be 21,666 LSBGs and 30,898
HSBGs by following  the same criteria given in Sect.2. Finally we
obtain about 280 LSBGs and 300 HSBGs with reliable observations
from UV to IR. If without considering IR, there are about 2100
LSBGs and 7300 HSBGs having multiwavelength observations from UV
to NIR. These will be good samples to study the multiwavelength
SEDs and stellar formation history of galaxies. One of the useful
model to analyze their SEDs is GRASIL (Silva et al. 1998).  This
model could produce the multiwavelength SEDs  of galaxies, such as
M51 (nearly face-on disk), M82 (starburst) and NGC 6090 (strongly
interacting galaxy) etc. Our sample could be compared with these
SEDs and hopefully they will be similar to those of M51 and/or M82.

In summary, we have selected a large sample of LSBGs from SDSS,
which could provide  interesting and important results for
understanding the properties of this kind of galaxies. Also it
must be very useful to better understand their contribution to the
local galaxy population. Their effects on the number density,
light density and light functions of galaxies in the local universe will be
further studied.

{\bf Acknowledgments:}
We thank the NSFC grant support under Nos. 10933001, 10973006,
10973015, 10673002, and the National Basic Research Program of
China (973 Program) Nos.2007CB815404,2007CB815406.


\begin{thebibliography}{}

\bibitem[Bothun et al. (1997)]{Bothun_etal97} 
{Bothun, G., Impey, C., McGaugh, S.} 1997, 
\textit{PASP}, 109, 745

\bibitem[Bruzual \& Charlot (2003)]{Bruzual_Charlot03}
{Bruzual, A. G. \& Charlot, S.} 2003, 
\textit{MNRAS}, 344, 1000

\bibitem[Chen et al. (2009)]{Chen_etal08} 
{Chen, X. Y., Liang, Y. C., Hammer, F., Zhao, Y. H., Zhong, G. H.} 2009, 
\textit{A\&A}, 495, 457

\bibitem[Cid Fernandes et al. (2005)]{cid_fernandes_etal05} 
 {Cid Fernandes, R., Mateus, A. \& Sodr\'{e}, L.} 2005b, 
 \textit{MNRAS}, 358, 363

\bibitem[Disney (1976)]{Disney76} 
{Disney, M.} 1976, 
\textit{Nature}, 263, 573

\bibitem[Freeman (1970)]{Freeman70}
{Freeman, K. C.} 1970, 
\textit{ApJ}, 160, 811

\bibitem[Impey \& Bothun (1997)]{ImpeyBothunb97}
{Impey, C., \& Bothun, G.} 1997, 
\textit{ARAA}, 35, 267

\bibitem[Impey et al. (1996)]{Impey_etal96}
{Impey, C. D., Sprayberry, D., Irwin, M. J., Bothun, G. D.} 1996, 
\textit{ApJS}, 105, 209

\bibitem[Kauffmann et al. (2003a)]{Kauffmann_etal03a}
{Kauffmann, G., Heckman, T. M., Tremonti, C. et al.} 2003a,
\textit{MNRAS}, 346, 1055

\bibitem[Kauffmann et al. (2003b)]{Kauffmann_etal03b}
{Kauffmann, G., Heckman, T. M., White, S. D. M. et al.} 2003b,
\textit{MNRAS}, 341, 33

\bibitem[Kniazev et al. (2004)]{Kniazev_etal04} 
{Kniazev, A. Y., Grebel, E. K., Pustilnik, S. A. et al.} 2004, 
\textit{AJ}, 127, 704

\bibitem[Liang et al. (2006)]{Liang_etal06} 
{Liang, Y. C., Yin, S. Y., Hammer, F., Deng, L. C., Flores, H., Zhang, B.} 2006, 
\textit{ApJ}, 652, 257

\bibitem[McGaugh (1996)]{McGaugh_96} 
{McGaugh S. S.} 1996, 
\textit{MNRAS}, 280, 337

\bibitem[Monnier-Ragaigne et al. (2003)]{Monnier-Ragaigne_etal.03}  
{Monnier-Ragaigne, D., van Driel, W., Schneider, S. et al.} 2003, 
\textit{A\&A}, 405, 99
    
\bibitem[O'Neil et al. (1997)]{O'Neil_etal97a} 
{O'Neil, K., Bothun, G. D., Cornell, M.} 1997a, 
\textit{AJ}, 113, 1212

\bibitem[Silva et al. (2003)]{Silva_etal03}
{Silva, L. et al.} 1998, 
\textit{ApJ}, 509, 103

\bibitem[Tremonti et al. (2004)]{Tremonti_etal04} 
{Tremonti, C. A.,  Heckman, T. M., Kauffmann, G. et al.} 2004, 
\textit{ApJ}, 613, 898
 
\bibitem[Zhong et al. (2008)]{Zhong_etal08} 
{Zhong, G. H., Liang, Y. C., Liu, F. S. et al.} 2008, 
\textit{MNRAS}, 391, 986

\end{thebibliography}
\end{document}